\documentclass[aps,prb,twocolumn,groupedaddress,draft,showpacs,intlimits,amsmath,amssymb,floats]{revtex4}
\newcommand{\bk}{\textbf{k}}
\newcommand{\bK}{\textbf{K}}
\newcommand{\bq}{\textbf{q}}

\newcommand{\bra}[1]{\langle #1|}
\newcommand{\ket}[1]{|#1\rangle}

\usepackage{bm}
\usepackage[final]{graphicx}
\usepackage{epsfig}

\begin{document}


\title{On the Question of Coincidence Between Energy Gaps and Kohn Anomalies} 


\author{S. Johnston$^{1,2}$}
\author{A. P. Sorini$^{2}$}
\author{B. Moritz$^{2,3}$}
\author{T. P. Devereaux$^{2,4}$}
\author{D. J. Scalapino$^{5}$}
\affiliation{$^1$IFW Dresden, P.O. Box 27 01 16, D-01171 Dresden, Germany}
\affiliation{$^2$Stanford Institute for Materials and Energy Science, 
SLAC National Accelerator Laboratory and Stanford University, Stanford, CA 94305, USA}
\affiliation{$^3$Department of Physics and Astrophysics, University of North Dakota, Grand Forks, ND 58202, USA}
\affiliation{$^4$Geballe Laboratory for Advanced Materials, Stanford University, Stanford, CA 94305, USA}
\affiliation{$^5$Department of Physics, University of California, Santa Barbara, CA 93106-9530 USA}
\begin{abstract}
Recently, neutron scattering spin echo measurements have provided high 
resolution data on the temperature dependence of the linewidth 
$\Gamma({\bf q},T)$ of acoustic phonons in conventional superconductors 
Pb and Nb.\cite{AynajianScience2008}  At low temperatures the merging 
of the $2\Delta(T)$ structure in the linewidth with a peak associated 
with a low lying $\hbar\omega_{\bf q_{KA}}$ Kohn anomaly suggested a 
coincidence between $2\Delta(0)$ and $\hbar\omega_{\bf q_{KA}}$ in Pb 
and Nb. Here we carry out a standard BCS calculation of the phonon 
linewidth to examine its temperature evolution and explore how 
close $2\Delta(0)/\hbar\omega_{\bf q_{KA}}$ must be to unity 
in order to be consistent with the neutron data.
\end{abstract}

\date{\today}
\maketitle

\section{Introduction}
Using resonant spin echo neutron scattering techniques, 
Aynajian {\it et al}.\cite{AynajianScience2008} have 
recently measured the linewidth of transverse acoustic phonons in 
high purity single crystals of Pb and Nb. At low temperatures, 
which are however above the superconducting transition temperature 
$T_c$, a plot of the phonon linewidth $\Gamma({\bf q}, T)$ as a 
function of the phonon wavevector ${\bf q}$ exhibits peaks which 
arise from Kohn anomalies.\cite{KohnPRL1957} When the temperature 
decreases below $T_c$ and the superconducting gap opens, one sees 
an expected decrease in the linewidth $\Gamma({\bf q}, T)$ for 
phonons having energy $\hbar\omega_{\bf q}$ less than twice the 
superconducting gap $\Delta(T)$. As $\hbar\omega_{\bf q}$ 
approaches $2\Delta(T)$, there is a rapid increase in $\Gamma({\bf q},T)$ 
associated with the peak in the quasiparticle density of states at the 
gap edge and the fact that the BCS coherence factor for a phonon to 
break a Cooper pair and decay into two quasiparticles approaches 
1 at threshold.\cite{BCS} However, as Aynajian {\it et al}. note, 
what is surprising is that as $T$ goes to zero, the feature in 
$\Gamma({\bf q},T)$ that is associated with $\hbar\omega_{\bf q}=2\Delta(T)$ 
appears to merge with a Kohn anomaly peak. This behavior is seen in both Pb 
and Nb, posing the question of why should the energy of a transverse 
acoustic phonon associated with a normal state Kohn anomaly coincide 
with twice the limit of the low temperature superconducting gap $2\Delta(0)$?

Motivated by this experimental result, we have carried out a 
standard BCS calculation of the temperature dependence of the 
transverse acoustic line width and examined what happens if $2\Delta(0)$ 
is near the energy associated with a normal state Kohn anomaly 
in $\Gamma({\bf q},T)$. In particular, we are interested in the 
evolution of $\Gamma({\bf q},T)$ as the temperature is lowered 
and $2\Delta(T)$ approaches the energy of the Kohn anomaly 
$\hbar\omega_{{\bf q}_{KA}}$.  How close to $\hbar\omega_{{\bf q}_{KA}}$ 
does the low temperature limit of $2\Delta(T)$ need to be for it to 
appear that the $2\Delta(0)$ structure in $\Gamma({\bf q}, T)$ 
merges with the Kohn anomaly structure as $T$ goes to zero?

\section{Formalism}

\begin{figure}[tr]
 \includegraphics[width=0.9\columnwidth]{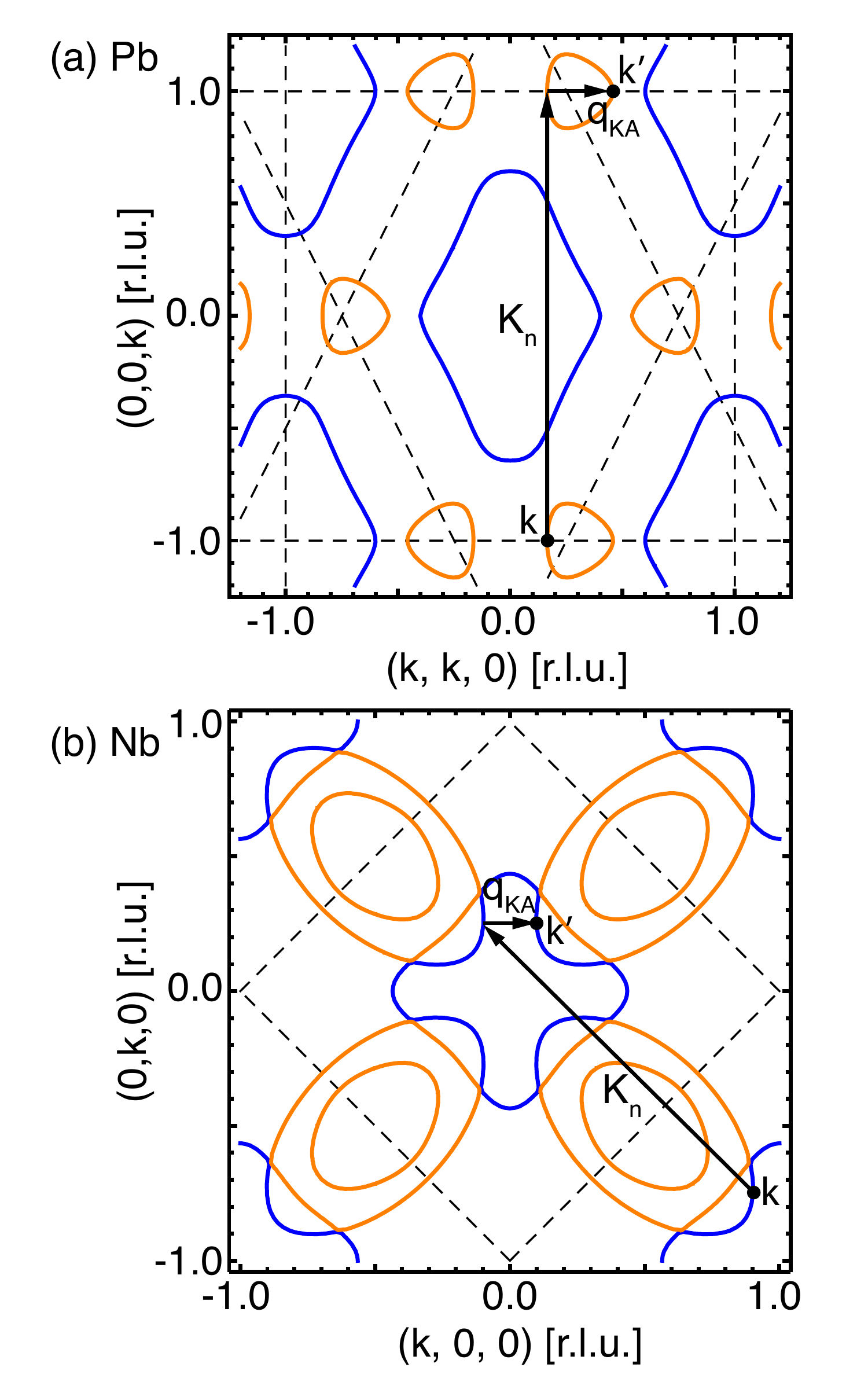}
 \caption{\label{Fig:FS} (Color online) The Fermi surfaces of 
 (a) Pb and (b) Nb.  The short black arrows indicate the $q_{KA}$ 
 wavevectors (r.l.u.) $q_{KA} =0.295$ in Pb and 
 $q_{KA} = 0.196$ in Nb.  The long arrow indicates the 
 reciprocal lattice vector ${\bf K}_n$ associated with the 
 Umklapp scattering process. 
 The location of the Bragg planes are indicated by thin dashed lines.}
\end{figure}

We begin by first examining the matrix elements for the electron   
coupling to the transverse acoustic modes. In clean materials, the coupling of the 
electrons to the low frequency transverse
phonons occurs through Umklapp scattering processes. 
As one knows, this is because the polarization
${\bf \hat\epsilon}_{\lambda}({\bf q})$ of a 
transverse phonon is orthogonal to ${\bf q}$. 
In Fig. \ref{Fig:FS}a,b we show Fermi surface sections for Pb and Nb, respectively, 
obtained from Density Functional Theory (DFT) calculations (ABINIT).\cite{Abinit}  
In the top figure for Pb, an Umklapp scattering process is shown in which 
an electron is scattered from $\bk$ to $\bk^\prime = \bk + {\bf K}_n + \bq$ 
with $\bq$ the wavevector of the transverse phonon and ${\bf K}_n$ a 
reciprocal lattice vector. 
In this case, the phonon wavevector ${\bf q}_{KA}$ that is shown connects two parts of the 
Fermi surface that have parallel tangents leading to a Kohn anomaly 
in the scattering rate and the phonon linewidth. A similar process for 
Nb is illustrated in the lower part of Fig. \ref{Fig:FS}. 

In the following calculations, we use an electron-phonon 
vertex $g_{\lambda}({\bf k,k^{\prime}})$ 
\begin{eqnarray}\label{Eq:Vertex}\nonumber
g_\lambda(\bk,\bk^\prime) = 
-i\sqrt{\frac{\hbar}{2MN\omega_\lambda(\bk-\bk^\prime)}} 
\hat{\epsilon}_\lambda(\bk-\bk^\prime) \cdot \\ \nonumber
\sum_{\bK_m,\bK_n}
(\bk + \bK_m - \bk^\prime - \bK_n) a^\dagger_{\bK_m}(\bk)a_{\bK_n}(\bk^\prime)
\times \\
\bra{\bk+\bK_m}U\ket{\bk^\prime+\bK_n}
\end{eqnarray}
for a transverse acoustic mode $\lambda$ which has a frequency $\hbar\omega_{\lambda}({\bf q})$ and a polarization vector ${\bf \hat\epsilon}_{\lambda}({\bf q})$. 
Here $M$ is the ion mass, $U$ is the lattice pseudo-potential, and $N$ 
is the number of lattice sites.  As discussed, for transverse phonons one needs 
an Umklapp process to couple the electrons to the direction of the ionic 
vibration given by $\hat{\epsilon}_{\lambda=T}$ such that 
$g_T \propto (\bK+\bq)\cdot \hat{\epsilon}_T = \bK\cdot\hat{\epsilon}_T$.  Then, as we will see, the linewidth
of the transverse phonons will exhibit a peak as ${\bf q}$ approaches the Kohn anomaly 
wavevector ${\bf q}_{KA}$.

\begin{figure}
 \includegraphics[width=\columnwidth]{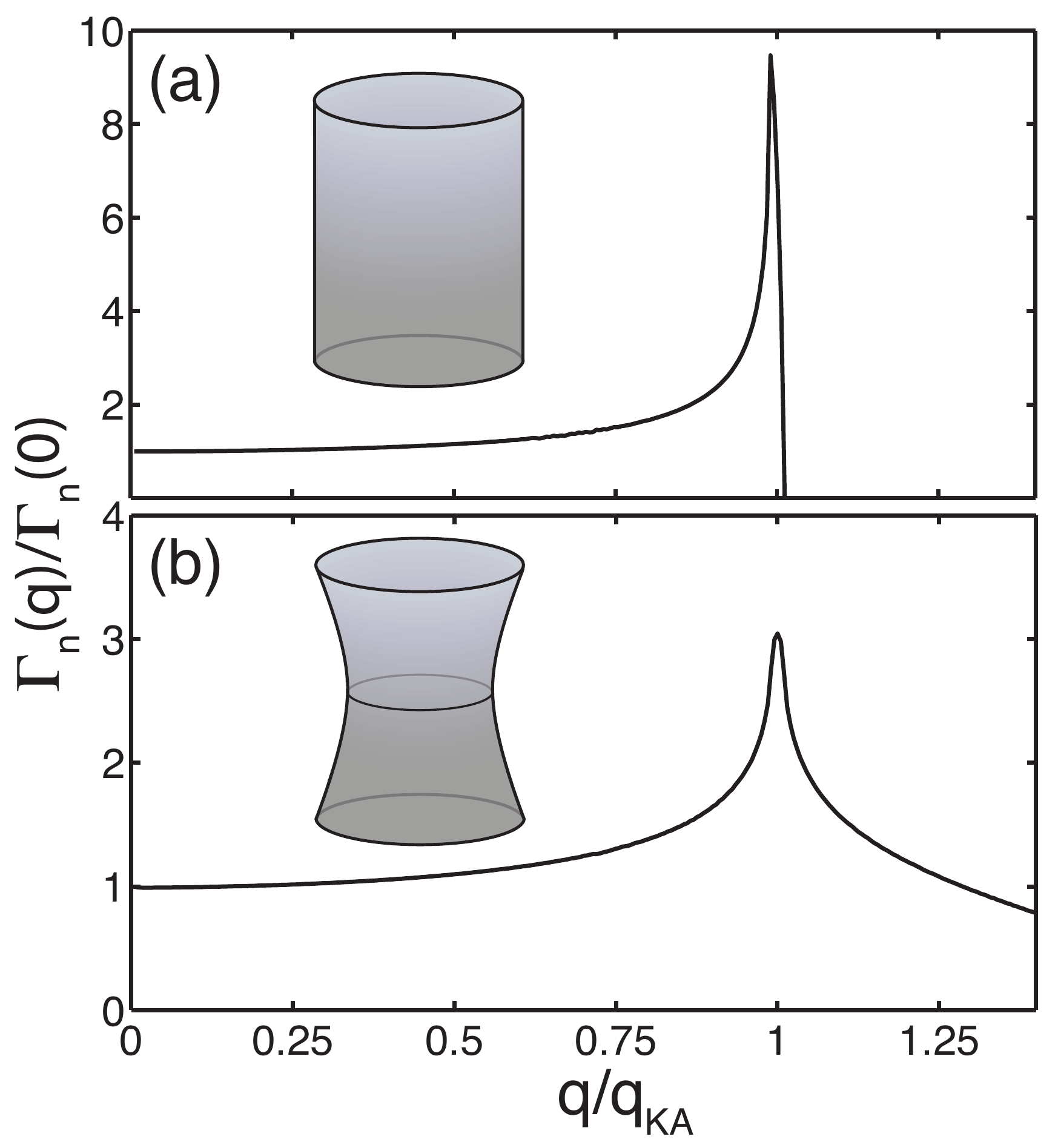}
 \caption{\label{Fig:gamma_ns} The zero temperature phonon linewidth  
 of the transverse acoustic branch in the normal state 
with $c_T/v_F = 0.01$ evaluated for 
 free electrons with (a) a cylindrical Fermi surface and (b) 
 concave Fermi surface.}
\end{figure}

To capture the essence of the Kohn-Umklapp scattering, we first consider 
the expression for the transverse acoustic phonon linewidth $\Gamma(\bq,T)$ 
in the normal state for the case in which the Fermi surface spanned by $\bq_{KA}$ 
is approximated by a cylinder of radius $k_F$ (Fig. \ref{Fig:gamma_ns}a).  In   
this case, $\Gamma(\bq,T)$ is given by
\begin{eqnarray}\label{Eq:NS} 
\Gamma(\bq,T)=\frac{\pi|g_\bK|^2}{N}\sum_\bk [f(\epsilon_\bk) - 
f(\epsilon_{\bk+\bq})] \nonumber\\
 \quad\quad \times\delta(\omega_\bq - \epsilon_{\bk+\bq}+\epsilon_{\bk})
\end{eqnarray}
with $f$ the Fermi factor and $\epsilon_\bk$ the 
electronic band dispersion. From here on we choose $\hbar=1$.  For simplicity, 
we have set $g_{\lambda=T}({\bf k,k^{\prime}})=g_{\bf K}$, the phonon mode 
energy $\omega_\bq = c_T|\bq|$, with $c_T$ the transverse speed of sound, 
and assumed a simple 2D free-electron dispersion $\epsilon_\bk = 
k^2/2m - \mu$.  Taking the $T = 0$ limit and making the change of 
variables $x = k/k_F$, Eq. (\ref{Eq:NS}) reduces to
\begin{eqnarray}
\Gamma(\bq,T=0)=\frac{mk_F|g_\bK|^2}{4\pi q}\int_0^1 xdx \int_0^{2\pi} d\phi 
\\ \nonumber
\times 
\left[\delta\left(\alpha_-(\bq)-x\cos(\phi)\right) - 
\delta\left(\alpha_+(\bq) - x\cos(\phi)\right) \right] 
\end{eqnarray} 
where $\alpha_\pm(\bq) = \frac{c_T}{v_F}\pm\frac{q}{2k_F}$ and  
$k_F$ and $v_F$ are the Fermi momentum and velocity, respectively.
After a little algebra we then obtain
\begin{eqnarray}\nonumber
\Gamma(\bq,T=0) = N_F|g_\bK|^2\frac{2k_F}{q} 
\bigg[ 
\sqrt{1-\alpha^2_-(\bq)}\Theta(1-\alpha_-^2(\bq))\\
-\sqrt{1-\alpha^2_+(\bq)}\Theta(1-\alpha_+^2(\bq))
\bigg]
\end{eqnarray}
where $N_F$ is the single-particle density of states per spin at the Fermi 
level and $\Theta(x)$ is the usual step function.  

$\Gamma(\bq,T=0)$ is plotted in Fig. \ref{Fig:gamma_ns}a for 
$c_T/v_F = 0.01$.  While the overall magnitude of the linewidth 
is determined by the ratio of $c_T/v_F$, the momentum dependence comes 
from simple phase space considerations.  One can see that the phonon 
linewidth grows rapidly for momentum transfers approaching $2k_F$ 
in this example and quickly falls to zero for larger momentum 
transfers as no phase space is available for scattering.  
This strong enhancement of the phonon linewidth in the normal state 
at $\bq$ corresponding to the Kohn anomaly will also be 
present in the superconducting state, with an additional 
kinematic constraint imposed by the breaking of Cooper pairs.  

\begin{figure}
 \includegraphics[width=\columnwidth]{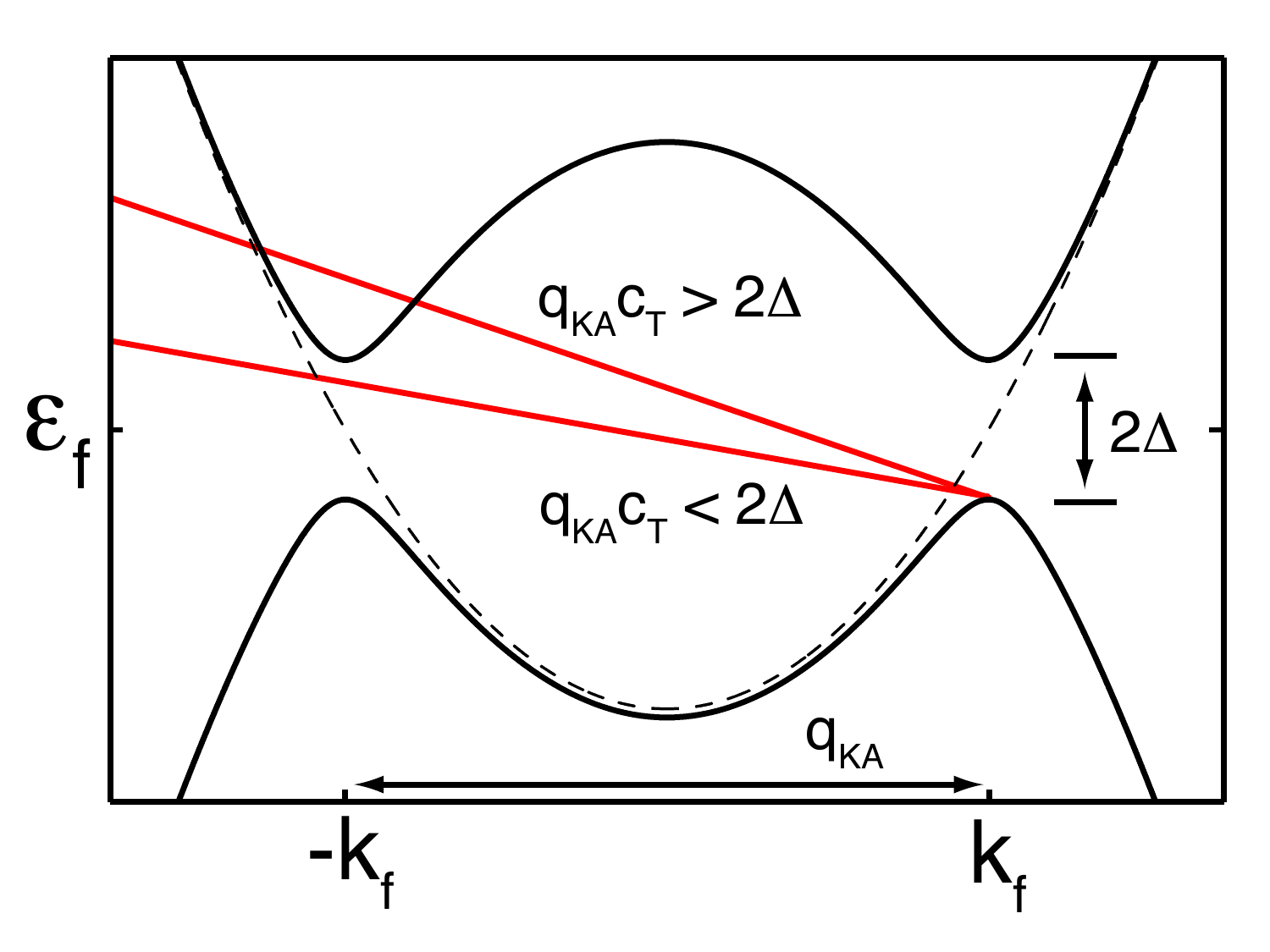}
 \caption{\label{Fig:banddiagram} (Color online) A schematic of the kinematic constraint 
 for the decay of an acoustic phonon in the superconducting state.}
\end{figure}

The kinematic constraint of phonon decay in the superconducting 
state brings the energy scale $2\Delta$ directly into play.  
This is shown in Fig. \ref{Fig:banddiagram}, 
which sketches quasiparticle scattering across the gap edge $2\Delta$. 
Due to the dispersion of the phonon, vertical scattering processes 
having no net wavevector transfers are kinematically forbidden. 
In order to bridge the gap the energy of the phonon must be at least 
$2\Delta$.  In other words, a finite wavevector transfer must 
occur where $q=2\Delta/c_T$. In addition, the dominant Kohn-Umklapp 
process $\bk^\prime - \bk = {\bf K}_n + \bq_{KA}$ 
involves a momentum transfer of $\bq_{KA}$, which in this sketch is 
$2k_F$.  Thus we have two conditions that lead to the conclusion 
that when $\omega(\bq)$ of the transverse acoustic phonon 
branch equals twice the superconducting gap, or in other words when 
$q = q_{KA} = 2\Delta/c_T$, an enhancement of the phonon decay 
will occur. 

\begin{widetext}
We next consider the transverse acoustic 
phonon decay rate in the superconducting state for this cylindrical model 
using conventional BSC 
theory.\cite{Parks}  The phonon self-energy can be obtained by evaluating 
the electron-hole bubble. 
In the superconducting state the phonon self-energy $\Pi(\bq,i\omega_m)$ 
is then given by 
\begin{equation}
\Pi(\bq,i\omega_m) = \frac{1}{N\beta} \mathrm{Tr}\sum_{n,\bk} 
|g_{\bf K}|^2 \hat{\tau}_3\hat{G}(\bk,i\omega_n) 
\hat{G}(\bk+\bq,i\omega_n+i\omega_m)\hat{\tau}_3
\end{equation}
where $\omega_n = (2n+1)\pi/\beta$ 
and $\omega_m = 2m\pi/\beta$ are Fermion and Boson Matsubara frequencies, 
$\mathrm{Tr}$ denotes the trace, and $\hat{G}$ is the 
electron propagator
\begin{equation}
\hat{G}(\bk,i\omega_n) = \frac{i\omega_n\hat{\tau}_0 + 
\epsilon_\bk\hat{\tau}_3 + \Delta_\bk\hat{\tau}_1}
{(i\omega_n)^2 - E_\bk^2} .
\end{equation}
Here $\hat{\tau}_i$ are the usual Pauli matrices and 
$E_\bk = \sqrt{\epsilon_\bk^2 + \Delta_\bk^2}$ is the quasiparticle 
energy.  After analytic continuation, the phonon self-energy is given by 
\begin{eqnarray}\label{Eq:gamma} \nonumber
\Pi(\bq,\omega_\bq)&=&\frac{1}{2N}\sum_\bk |g_{\bf K}|^2 
\bigg\{A_+(\bk,\bq)[f(E_\bk) - f(E_{\bk+\bq})] 
\bigg[ 
\frac{1}{\hbar\omega_\bq - E_\bk + E_{\bk+\bq} + i\delta} -
\frac{1}{\hbar\omega_\bq + E_\bk - E_{\bk+\bq} + i\delta}\bigg] 
\\
&+&A_-(\bk,\bq)[f(-E_\bk) - f(E_{\bk+\bq})] 
\bigg[\frac{1}{\hbar\omega_\bq + E_\bk + E_{\bk+\bq} + i\delta} -
\frac{1}{\hbar\omega_\bq - E_\bk - E_{\bk+\bq} + i\delta}\bigg] 
\bigg\}
\end{eqnarray}
with the coherence factors defined as 
\begin{equation}
A_\pm(\bk,\bq) = 1 \pm 
\frac{\epsilon_\bk\epsilon_{\bk+\bq} - \Delta_\bk\Delta_{\bk+\bq}}
{E_\bk E_{\bk+\bq}}.
\end{equation}  
The $\bq$-dependent phonon linewidth $\Gamma(\bq,T)$ is then determined from the 
imaginary part of $\Pi(\bq,\omega_{{\bf q}})$. 
\end{widetext}

The first two terms in Eq. \ref{Eq:gamma} describe quasiparticle scattering processes. For these processes, the BCS coherence factor $A_+$ vanishes at the threshold and this, along with the depletion of the thermal quasiparticle populations as the gap opens, suppresses their contribution to the phonon linewidth. The fourth term in Eq. \ref{Eq:gamma} corresponds to a process in which a phonon breaks a pair, creating two quasiparticles with wavevectors ${\bf k+q}$ and $-{\bf k}$. This requires that the phonon energy $\omega_{\bf q}$ be greater or equal to $2\Delta(T)$. In this case, the BCS coherence factor $A_-$ goes to 1 at threshold where $E({\bf k+q})=E({\bf k})=\Delta(T)$ and there is a sudden increase in the linewidth.

\begin{figure}
 \includegraphics[width=\columnwidth]{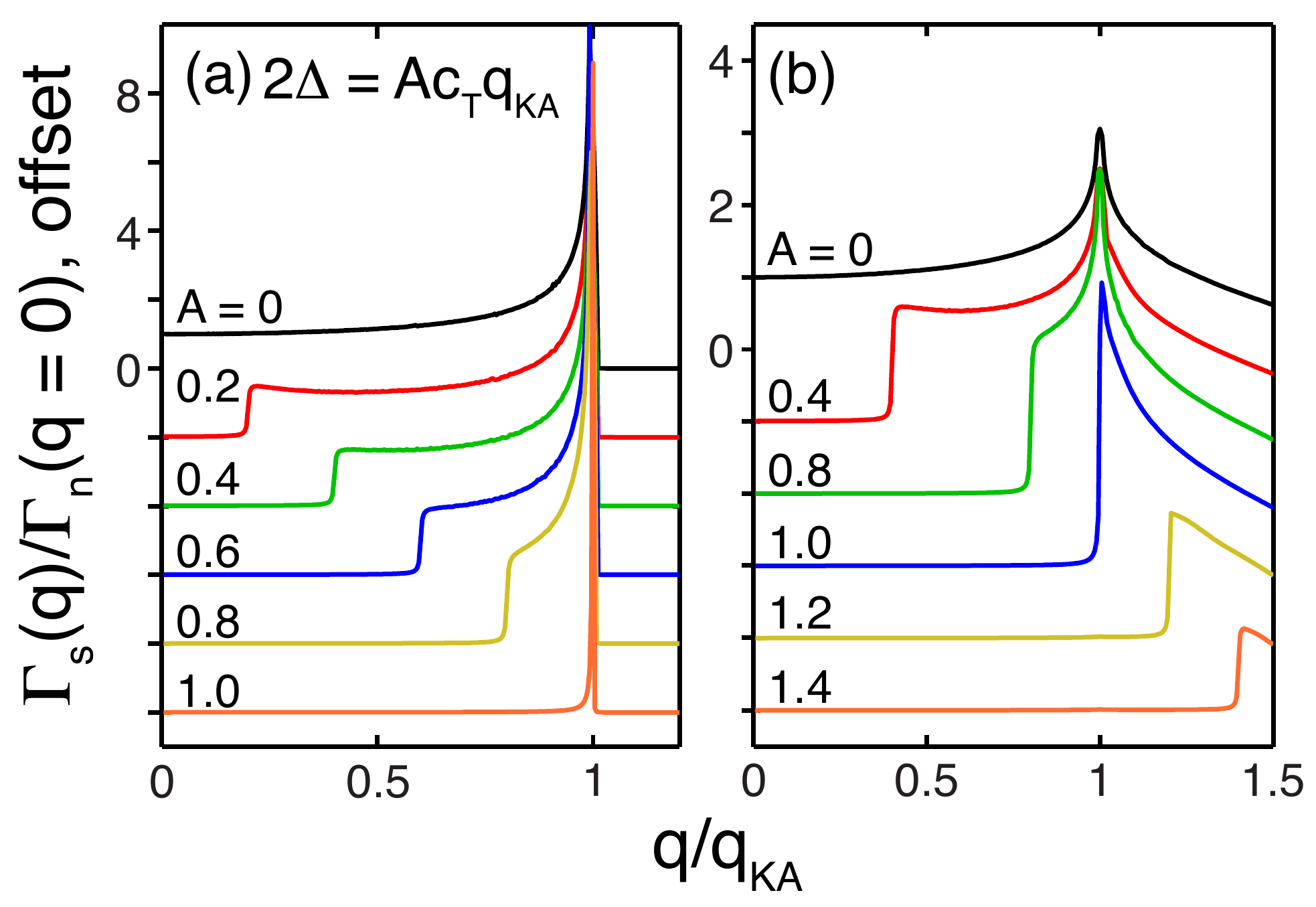}
 \caption{\label{Fig:gamma_sc} 
 (Color online) The normalized transverse acoustic phonon 
 linewidth in the superconducting state at $T=0$ for various values 
 of the superconducting gap $2\Delta = Ac_Tq_{KA}$.  
 (a) $\Gamma(q)$ for 
 a perfectly cylindrical Fermi surface.  (b) $\Gamma(q)$ for a 
 cylindrical Fermi surface with a concave warping along the $k_z$ 
 direction (see text).
 }
\end{figure}

Before turning to the results for the linewidth in Pb we first 
consider two simplified cases at $T=0$, shown in Fig. \ref{Fig:gamma_sc}. 
Here we have set the phonon energy $\omega_{\bf q}=c_Tq$ and plotted $\Gamma({\bf q},T=0)$ 
versus $q/q_{KA}$.
The $\Delta=0$ curve is identical to Fig. \ref{Fig:gamma_ns}. 
As the superconducting 
gap opens $\Gamma({\bf q},T=0)$ is suppressed for $c_Tq < 2\Delta(0)$ due 
to the loss of phase space for electron-phonon scattering.  
This produces an onset (or ``knee")  
in $\Gamma({\bf q},T=0)$ at an energy corresponding to the gap edge. 
Note that in this case one expects a knee rather than 
a peak because $q\xi \sim v_F/c_T \gg 1$.\cite{KleinDierker} 
(The knee is also  
somewhat smeared here due to the finite broadening $\delta = c_T/40$ used.)

The height of the onset is controlled by the momentum $q$ for breaking 
a Cooper pair into two quasiparticles carrying momenta ${\bf k}$ 
and ${\bf k-q}$ respectively.\cite{RMP,KleinDierker}
As $\Delta$ is made larger, the onset at $c_Tq=2\Delta$ associated 
with pair-breaking in the superconducting state moves out 
towards the Kohn anomaly at $q_{KA}$.  For a cylindrical Fermi surface, the Kohn anomaly occurs at $q_{KA}=2k_F$ and when $2\Delta(0)=c_Tq_{KA}$, the pair-breaking onset  coincides with the Kohn anomaly peak. If $2\Delta(0)$ exceeds $c_Tq_{KA}$, 
the Kohn anomaly peak is suppressed by kinematics as the energy to 
break a pair is greater than $c_Tq_{KA}$.

For a cylindrical Fermi surface with $2\Delta(0) > c_Tq_{KA}$, 
$\Gamma(q)$ is suppressed 
due to phase space considerations previously discussed for the normal state. 
However, if the Fermi surface has some degree of curvature along the $k_z$ 
direction such a sharp cut-off will not occur.  
To illustrate this, in Fig. \ref{Fig:gamma_sc}b 
we plot $\Gamma(\bq,T=0)$ for a Fermi surface that has a 
concave warping along the $z$ direction (see Fig. \ref{Fig:gamma_ns}b). 
The electronic band dispersion 
has again been modeled by a free electron dispersion but with 
$m_x = m_y = m$ and $m_z = -5m$.\cite{Footnote} 
For such a dispersion $q_{KA}$ corresponds to the spanning condition 
across the narrowest portion of the Fermi surface ($k_z = 0$).  
As can be seen in Fig. \ref{Fig:gamma_sc}b, the concave curvature of the 
Fermi surface provides phase space for scattering with momentum transfers  
$\bq > \bq_{KA}$ and the sharp cutoff in $\Gamma({\bf q},T=0)$ is no longer present. 
With the opening of the superconducting gap, $\Gamma({\bf q},T=0)$ is suppressed 
for $c_Tq < 2\Delta(0)$, just as in the previous case.  For 
$2\Delta(0) = c_Tq_{KA}$ a remnant of the Umklapp-Kohn peak 
remains. As the gap is increased further ($2\Delta(0)> c_Tq_{KA}$) 
the phase space associated with the Kohn peak is gapped out and 
the peak in $\Gamma({\bf q},T=0)$ is thus suppressed.  

\section{Results for Lead}
With the simple examples of the previous sections we are now 
ready to turn to the phonon linewidth in Pb. 
To obtain the electron dispersion the DFT bandstructure for Pb 
was calculated on a regular grid of $100\times100\times100$ 
momentum points per quadrant 
of the first Brillouin zone and a linear interpolant 
was used to obtain energies at intermediate momenta.   
For the phonon dispersion we again assume a linear phonon 
dispersion $\omega(\bq) = c_T|q|$, with $c_T = 7.9$ meV/[r.l.u.].  
The transition temperature $T_c = 7.2$ K sets the temperature scale 
and we use an intrinsic broadening $\delta = 0.01$ meV throughout.   
Finally, we note that an explicit evaluation of the matrix element for 
Umklapp scattering $g_{\bk,\bk^\prime}$ given by Eq. (\ref{Eq:Vertex}) adds an 
additional level of difficulty to the problem.  Therefore, for 
simplicity, we approximate the matrix element with a constant $g_{\bf K}$ 
and restrict the momentum sum to the region near the 
orange (light) Fermi surfaces shown in Fig. \ref{Fig:FS}a.   

\begin{figure}
 \includegraphics[width=\columnwidth]{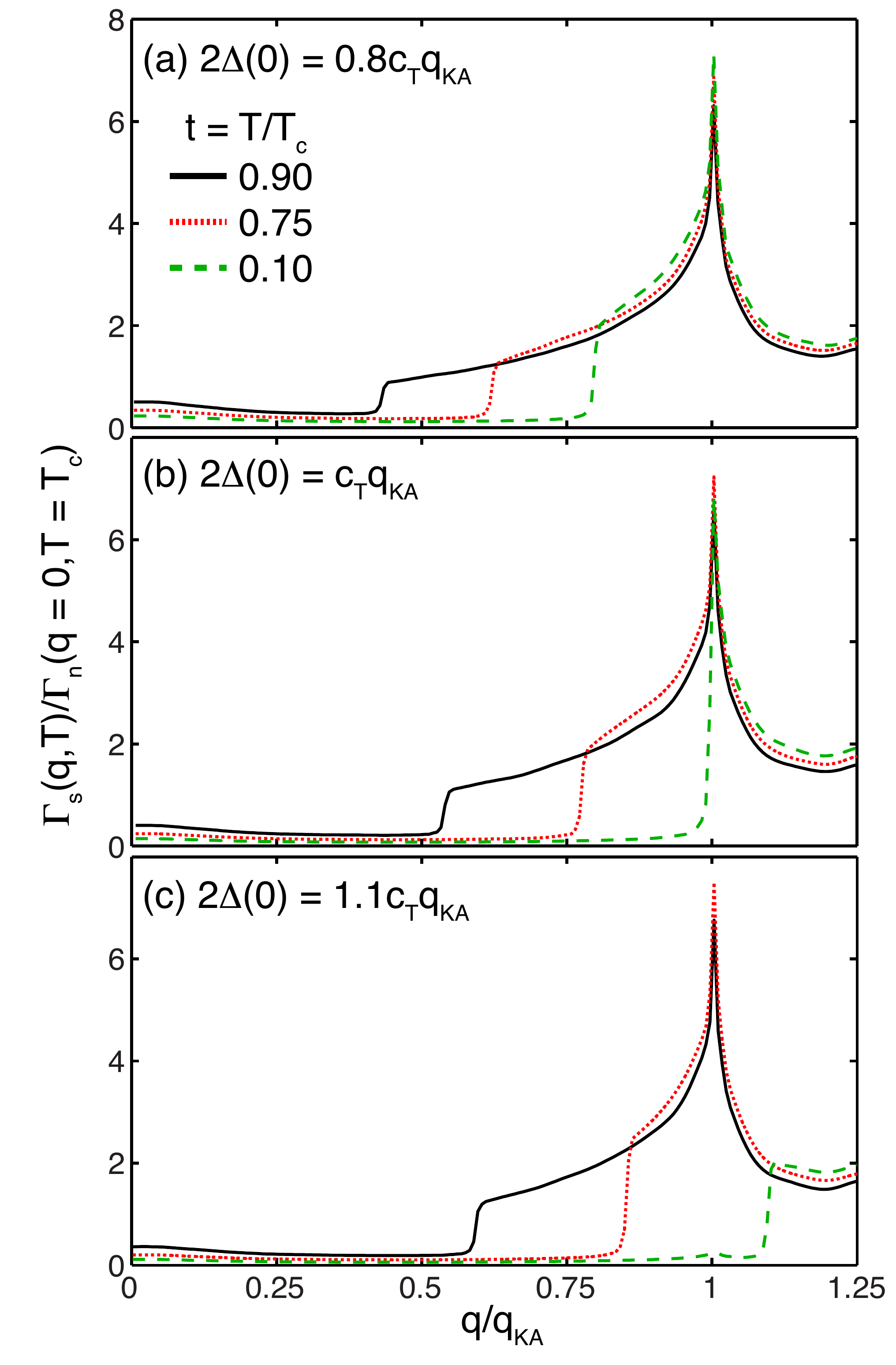}
 \caption{\label{Fig:Lead} 
 (Color online) The linewidth of a 
 transverse acoustic phonon in Pb as a function of 
 reduced temperature $t=T/T_c$ for various values of 
 $2\Delta(T=0)$ as indicated. 
 }
\end{figure} 

The results are shown in Fig. \ref{Fig:Lead} as a function of 
temperature for gap sizes ranging from 
$2\Delta(T=0) = 0.8 c_Tq_{KA}$ to $1.1 c_Tq_{KA}$.  
The qualitative behavior of $\Gamma({\bf q},T)$ is similar to that 
which was found for the simplified models considered in the previous section. 
Above T$_c$ the phonon linewidth is finite for all values of $\bq = (q,q,0)$ and 
has a peak at $q = q_{KA} = 0.285$ [r.l.u.], which is associated 
with the Kohn anomaly indicated in Fig. \ref{Fig:FS}. 
As the temperature is lowered across T$_c$, the gap opens following an assumed 
BCS temperature dependence.  For $c_Tq < 2\Delta(T)$, $\Gamma(q,T)$  
is suppressed and the expected $2\Delta$ onset (knee) forms.  
(Here, $\Gamma(\bq)$ has a finite value for  
$c_Tq < 2\Delta(T)$ which is exponentially suppressed as $T$ is lowered. 
This is due to the non-zero contributions of the 
first two terms in Eq. (\ref{Eq:gamma}) and corresponds to the thermal 
occupation of quasiparticle states across the gap edge.)  As $T$ is lowered 
further $2\Delta(T)$ grows and the knee in $\Gamma(q,T)$ moves 
towards the Kohn peak.  If $2\Delta(0)$ is smaller 
than $c_Tq_{KA}$ this knee stops short of the peak at the lowest 
temperatures (Figs. \ref{Fig:Lead}a) while for 
for $2\Delta(0) = c_Tq_{KA}$ it merges with the peak 
(Fig. \ref{Fig:Lead}b).  
Finally, if $2\Delta(0) > c_Tq_{KA}$ (Fig. \ref{Fig:Lead}c), then for sufficiently 
low temperatures the Kohn peak is suppressed similar to the results shown in Fig. 4b.

Thus within a BCS framework, $\Gamma({\bf q},T)$ depends upon 
the shape of the Fermi surface and ${\bf q}_{KA}$, the velocity 
of sound for the transverse acoustic branch, and the 
magnitude of the superconducting gap.
The appropriate parameters for Nb and Pb are summarized in  
Tbl. \ref{Tbl:parameters}.  For Pb we estimate $2\Delta(0) = 0.95c_Tq_{KA}$, 
which corresponds closest to Fig. \ref{Fig:Lead}b,  
while for Nb we estimate $2\Delta(0) = 0.8 c_Tq_{KA}$ corresponding 
to Fig. \ref{Fig:Lead}a.   

Comparing our results to Figs. 3 and 4 of Ref. 
\onlinecite{AynajianScience2008} we find that agreement 
with the experimental data for Pb is good while the 
agreement for the case of Nb is less clear. 
For Pb we find $2\Delta(0) \sim 0.95c_Tq_{KA}$ and we therefore 
expect a knee to form in $\Gamma({\bf q},T)$ which tracks out to 
the Kohn peak as the temperature is lowered.  
This behavior is similar to what is observed experimentally 
(Fig. 3 of Ref. \onlinecite{AynajianScience2008}).  In the case of Nb 
$2\Delta(0) \sim 0.80 c_Tq_{KA}$ and we therefore expect the knee 
to approach the Kohn peak but stopping short at the lowest 
temperatures leaving a pronounced knee in the 
observed linewidth.  Examining Fig. 4b of 
Ref. \onlinecite{AynajianScience2008} it is difficult to 
determine if such a knee is present in the data.  
Finally, we note that our calculations predict that 
the Kohn peak should be suppressed when $2\Delta(0) > c_Tq_{KA}$.  Therefore, one 
clear way to test the conclusions of this work would be to examine the linewidth of 
the transverse acoustic branch in a material where $c_Tq_{KA}<2\Delta(0)$.

\begin{table}[tr]
\begin{center}
\begin{tabular}{|l|| c|c|c|c| r|}
\hline
& $2\Delta(0)$ [meV] & $c_T$ [meV/r.l.u.] & $q_{KA}$ [$r.l.u.$] & $2\Delta(0)/c_Tq_{KA}$ \\\hline
Pb & 2.70 & 7.93 & 0.36 & 0.95 \\
Nb & 3.06 & 21.3 & 0.18 & 0.80 \\
\hline
\end{tabular}
\caption{\label{Tbl:parameters}
The relevant parameters for the elemental superconductors Pb and Nb. 
The values for Pb have been estimated from 
Ref. \onlinecite{AynajianScience2008}. The gap for Nb was 
obtained from Ref. \onlinecite{Carbotte}. The
transverse speed of sound in Nb was obtained from Ref. \onlinecite{Holt}.}
\end{center}
\end{table} 

\section{Conclusions}
We have seen that the momentum and temperature dependence 
of the transverse acoustic phonon linewidth $\Gamma({\bf q},T)$ 
in the superconducting state depends on $\omega_{{\bf q}_{KA}}$ 
and $2\Delta(T)$. While both of these energies depend upon the 
bandstructure and phonon dispersion, there is nothing that 
should lock them together in the traditional theory. Thus 
while it is known that the Kohn anomaly wavevector ${\bf q}_{KA}$ 
shown in Fig. 1 gives rise to a small kink in $\alpha^2F(\omega)$ 
associated with the energy $\omega_{{\bf q}_{KA}}$ at which the 
transverse phonon begins to contribute to the pairing interaction, 
this is a small feature and plays no role in determining 
$\Delta(0)$.\cite{Parks} Therefore, within the BCS framework, 
the fact that $2\Delta(0)$ is close to the energy of a Kohn 
anomaly $\omega_{{\bf q}_{KA}}$, must be viewed as a 
coincidence. Furthermore, as noted, the fact that the 
wavelength of the phonon is small compared to the coherence 
length leads to a knee-like feature  at $2\Delta(0)$ 
rather than a peak. Therefore if $\omega_{{\bf q}_{KA}} > 2\Delta(0)$, 
the Kohn anomaly remains as the dominant feature at low temperatures. 
However, as shown in Fig. 4b and 5a, if 
$2\Delta(0)$ is slightly less than the Kohn anomaly phonon 
energy $\omega_{{\bf q}_{KA}}$, the $2\Delta(T)$ structure 
can appear to merge with the Kohn anomaly peak in 
$\Gamma({\bf q},T)$ as $T$ goes to zero. Thus we would 
conclude that it is an interesting coincidence that 
$2\Delta(0)$ is only slightly smaller than the energies 
of the Kohn anomalies in both Pb and Nb, but it 
does not necessarily mean that the superconducting gap is 
determined by the Kohn anomaly itself and does not 
force $2\Delta(0)=\omega_{{\bf q}_{KA}}$.

\section{Acknowledgements}
We would like to acknowledge B. Keimer for useful 
discussions. This work was supported by the US Department of 
Energy, Office of Basic Energy Sciences under contract No DE-AC02-76SF00515. 
D. J. S. acknowledges the Center for Nanophase Materials Science, which 
is sponsored at Oak Ridge National Laboratory by the Division of 
Scientific User Facilities, U.S. Department of Energy and thanks the 
Stanford Institute of Theoretical Physics for their hospitality.  
S. J. would like to acknowledge financial support from 
the Natural Sciences and Engineering Research Council of Canada 
and the Foundation for Fundamental Research on Matter. S. J., B. M. 
and T. P. D. would also like to thank the Walther Meissner Institute 
for their hospitality during the writing of this manuscript.


\begin{thebibliography}{99}
\bibitem{Footnote}
The values of the mass will affect the structure in $\Gamma(q)$ 
for $q > q_{KA}$ and specifically how quickly it falls off at large 
$\bq$. It does not change the qualitiative features arising from the opening 
of the gap which we wish to discuss here.
\end{thebibliography}
\end{document}